\documentclass[10pt]{article}
\usepackage{amssymb}
\usepackage{latexsym}
\usepackage{amsmath}
\newcommand{\x}{\mbox{\rm{\bf \^ x}}}
\newcommand{\p}{\mbox{\rm{\bf \^ p}}}
\newcommand{\z}{\mbox{\rm{\bf \^ z}}}
\newcommand{\bp}{\mbox{\rm{\bf p}}}
\newcommand{\bz}{\mbox{\rm{\bf z}}}

\newcommand{\xj}{\mbox{\rm{\bf \^ x}}_j}
\newcommand{\xk}{\mbox{\rm{\bf \^ x}}_k}
\newcommand{\pk}{\mbox{\rm{\bf \^ p}}_k}
\newcommand{\zk}{\mbox{\rm{\bf \^ z}}_k}
\newcommand{\zj}{\mbox{\rm{\bf \^ z}}_j}
\newcommand{\pj}{\mbox{\rm{\bf \^ p}}_j}
\newcommand{\A}{\mbox{\rm{\bf \^ A}}}
\newcommand{\Ad}{\bf \mbox{\rm{\A}}^\dagger}
\newcommand{\f}{f({\p})}
\newcommand{\xl}{\x_{\lambda}}
\newcommand{\et}{\mbox{\rm{and}}}

\newcommand{\cD}{{\cal D}}
\newcommand{\cU}{{\cal U}}
\newcommand{\cZ}{{\cal Z}}
\newcommand{\cW}{{\cal W}}
\newcommand{\cL}{{\cal L}}
\newcommand{\cB}{{\cal B}}
\newcommand{\xp}{i\,f(p)\partial_p}
\newcommand{\ZZ}{\mathbb Z}
\newcommand{\RR}{\mathbb R}
\newcommand{\NN}{\mathbb N}
\newcommand{\Id}{\mbox{\rm \bf 1\hspace{-1.2 mm}I}}%OPERATIONS

\newcommand{\KMM}{K. M. M. }
\begin{document}
\begin{flushright} UMH--MG--00/02 \\\today\\ \end{flushright}
\vspace{.1cm}
\begin{center} {\large About maximally localized states in quantum mechanics}\\
\vspace{.5cm} S.~Detournay, C.~Gabriel\footnote{Postdoctoral
researcher of the Belgium National Fund for Scientific
Research.}$^,$\footnote{ E-mail : claude.gabriel@umh.ac.be} and
Ph.~Spindel\footnote{ E-mail :
philippe.spindel@umh.ac.be}\\
\vspace{.2cm} {\it M\'ecanique et Gravitation}\\ {\it Universit\'e
de Mons-Hainaut, 20 Place du Parc}\\ {\it 7000 Mons, Belgium}\\
\end{center} \vspace{.1cm}

\begin{abstract}
We analyze the emergence of a minimal length for a large class of
generalized commutation relations, preserving commutation of the
position operators and translation invariance as well as rotation
invariance (in dimension higher than one). We show that the
construction of the maximally localized states based on squeezed
states generally fails. Rather, one must resort to a constrained
variational principle.
\end{abstract}
{\small {\bf Keywords: }{maximally localized states, squeezed
state,
variational principle}\\
{\bf PACS: }{03.65.Ca,03.65.Db}}

\section{Introduction}
In a series of articles (\cite{K1}-\cite{K3}), Kempf and
collaborators have studied the quantum mechanical implications of
a modification of the usual canonical commutation relations which
is inspired by noncommutative geometry \cite{K3} and string theory
(see for instance refs \cite{stref}). Assuming that
$[\x,\p]=if(\x,\p)$ these authors show that a minimal length
appears naturally from the Heisenberg uncertainty relation
\begin{equation}
\Delta \x\,\Delta\p \geq\frac12|\langle[\x,\p]\rangle|.
\label{Heis}
\end{equation}
This crucial modification of fundamental commutators is physically
grounded, and lies at the basis of recent theoretical models
aiming at describing quantum geometry or quantum gravity. The
emergence of a minimal localization length seems indeed to be a
recurrent feature when trying to approach quantum gravity and has
far reaching consequences (see e.g. \cite{Garay} and references
therein). This short distance structure could for instance allow
to get rid of ultraviolet divergencies (see \cite{KeMa} for a
discussion in the context of modified commutation relations). The
introduction of a naturel ultraviolet cutoff could also affect the
spectrum of the cosmological microwave background \cite{K4,K5}. We
will focus in this note on the physical states displaying the
minimal length uncertainty, the so-called {\itshape maximally
localized states}. In reference \cite{K2} , Kempf, Mangano and
Mann (\KMM hereafter) claim to obtain such states as particular
squeezed states. We shall show that it is only in a few special
cases that maximally localized states can be obtained as squeezed
states but that, in general, it is a variational principle that
leads to more satisfactory results. However we wish to emphasize
that our work only aims to provide what we believe to be a better
definition of maximally localized states, but the credit for the
idea that a modification of the canonical commutation relation
leads to a fundamental minimal length belongs to the authors of
refs \cite{K1}-\cite{K3}.

Our paper is organized as follows. We restrict ourselves to space
translation invariant situation {\it i.e.} we assume the
commutator to depend only on $\p$. In section {\bf 2} we analyze
the \KMM construction and display its weaknesses. Section {\bf 3}
consists in a discussion of a definition of physical maximally
localized states, based on a minimization principle of the spread
in space of the states (in one dimension). Section {\bf 4}
illustrates some physical aspects of our analysis through study of
the quantization of the particle in a box as well as the
asymptotic energy eigenvalues of the harmonic oscillator,
independently of the details of the commutation relation. In
section {\bf 5} we show that the extension to translation
invariant modified commutation relations in an arbitrary number of
dimensions, under the assumption of commuting position operators,
can be completely described by the use of a single function.
Moreover assuming rotational invariance, we then compare the
maximally localized states defined by our variational principle to
those proposed by \KMM and show that ours lead to a finer spatial
localization. We conclude our work in section {\bf 6} by a brief
summary of our results.
\section{\KMM construction} Hereafter
we assume the operator function $f$ to be only function of $\p$;
this insures the space translation invariance of the commutation
relation. In $p$-representation, the function $f(p)$ has to be
real. Moreover we assume it to be everywhere greater or equal to a
strictly positive constant
\begin{equation}
\exists\, C>0\,,\ \forall p\in \RR\ :\ f(p)>C\label{C1}
\end{equation}
{\it i.e.} whatever is the state, the operators $\x$ and $\p$
acting on it, never commute. In a first step let us restrict
ourselves to the one dimensional problem
\begin{equation} [\x,\p]=i \f\qquad .
\end{equation} In a $p$-representation these operators are realized as
\begin{equation}
\x=i\,f(p)\,\partial_p\qquad \et \qquad \p=p \in \RR\qquad .
\end{equation} \KMM construction of maximally localized state
consists in taking among the squeezed states, {\it i.e.} states
that satisfy the relation
\begin{equation}\label{Hs}
\Delta \x\,\Delta\p = \frac12|\langle[\x,\p]\rangle|\qquad ,
\end{equation} the one leading to the minimal value of $\Delta \x$.
Squeezed states depend on three parameters, denoted $\xi$, $\eta$
and $\Lambda$, and are the eigenstates $|\Psi_\Lambda\rangle$,
with vanishing eigenvalue, of the operators $\A_\Lambda$ defined
as
\begin{eqnarray}
\A_\Lambda|\Psi_\Lambda\rangle\equiv\left[(\x-\xi)+i\Lambda(\p-\eta)
\right)|\Psi_\Lambda\rangle=0\label{CS}
\end{eqnarray}
{\it i.e.} in $p$-representation
\begin{equation}
\langle p|\Psi_\Lambda\rangle \propto
\exp[(\Lambda\,\eta-i\,\xi)z(p)-\Lambda\, u(p)]\label{sqstate}
\end{equation}
where
\begin{equation}
z(p)=\int_0^pf^{-1}(q)\,dq\qquad{\rm and}\qquad
u(p)=\int_0^pq\,f^{-1}(q)\,dq\qquad .\label{zu}
\end{equation}
These states are always normalizable when the values of $z(\pm
\infty)$ are finite, otherwise $\Lambda$ must be positive, and it
is easy to check that the parameters $\xi$ and $\eta$ correspond
to the expectation values of the operators $\x$ and $\p$ if and
only if
\begin{equation}
\Lambda>0\qquad {\rm and}\qquad
\lim_{p\to\pm\infty}u(p)=\infty\qquad . \label{as1}
\end{equation}
Now, let us compute the norm of the state
$\A_\ell|\Psi_\Lambda\rangle$. We obtain
\begin{eqnarray}
\langle \Psi_\Lambda\A_\ell|\A_\ell \Psi_\Lambda\rangle
&=&(\ell-\Lambda)^2 \langle\Psi_\Lambda|(\p
-\eta)^2|\Psi_\Lambda\rangle\label{ellL}\\
&\geq&0
\end{eqnarray}
and moreover, if (and only if) $u(|p|)$ grows quickly enough at
infinity so that we have
\begin{equation}
\lim_{p\to\pm\infty}p\exp[2\,\Lambda ( \eta \,z(p)-
u(p))]=0\qquad.\label{as2}
\end{equation}
we also deduce, after an integration by parts, that:
\begin{eqnarray}\label{AA1}
\langle \Psi_\Lambda\A_\ell|\A_\ell \Psi_\Lambda\rangle
&=&\langle\Psi_\Lambda|\Ad_\ell\A_\ell\Psi_\Lambda\rangle\\ &=&
\langle\Psi_\Lambda| (\x-\xi)^2-\ell\,f(\p)+\ell^2
(\p-\eta)^2|\Psi_\Lambda\rangle\qquad . \label{AA2}
\end{eqnarray}

Introducing $(\Delta
\x)_\Lambda=\sqrt{\langle\Psi_\Lambda|\x^2|\Psi_\Lambda\rangle/\langle\Psi_\Lambda|\Psi_\Lambda\rangle
- \xi^2}$ and a similar definition for $(\Delta \p)^2_\Lambda$, we
obtain by identifying eq.(\ref{ellL}) with eq. (\ref{AA2}) that:
\begin{eqnarray}
(\Delta\x)_\Lambda^2(\Delta\p)_\Lambda^2 &=&\frac
14\langle f(\p)\rangle_\Lambda^2\qquad\qquad ,\\
\Lambda &=& \frac{\langle
f(\p)\rangle_\Lambda}{2(\Delta\p)_\Lambda^2}\qquad\ \,\qquad,\\
(\Delta\x)_\Lambda^2&=&\frac 14 \frac{\langle
f(\p)\rangle_\Lambda^2}{(\Delta\p)_\Lambda^2}
=\Lambda^2(\Delta\p)_\Lambda^2\qquad .
\end{eqnarray}
Accordingly, if ${\langle
f(\p)\rangle_\Lambda^2}/{(\Delta\p)_\Lambda^2}$ admits a strictly
positive lower bound, we deduce the existence of a minimal
uncertainty on the position operator {\it on the subset of
squeezed states}. The evaluation {\it a priori} of this minimum is
not simple except in two cases: if the mean values $\langle
f(\p)\rangle_\Lambda$ are bounded, in which case the minimum is
zero, or if $f(\p) = 1+ \beta^2\p^2$, the case discussed in detail
by \KMM \cite{K2}. To avoid unnecessary clutter, in the following
we shall made use of units such that constants, like $\beta$ in
the previous equation, are set equal to unity. When the previous
assumptions (\ref{as1}, \ref{as2}) are verified, in the
$p$-representation we may obtain a lower bound for the space
uncertainty of the squeezed states by calculating the minimum
(with respect to $\Lambda$ and $\eta$) of the integral expression
of $(\Delta \x)_{\Lambda}$:
\begin{equation} (\Delta
\x)^2_{\Lambda}=\Lambda^2\,\frac{\int_{-\infty}^{+\infty}(p^2-\eta^2)\exp[-2\Lambda
\,(u(p)-\eta\,z(p))]f^{-1}(p)\,dp}{\int_{-\infty}^{+\infty}\exp[-2\Lambda
\,(u(p)-\eta\,z(p))]f^{-1}(p)\,dp}\label{dxLdp}\qquad .
\end{equation}
In particular, under our assumption that for large values of
$\vert p \vert$, the function $f(p)$ behaves as $\vert
p\vert^{1+\nu}$ --- with $-1\leq\nu\leq 1$ in order to admit a
strictly positive lower bound and to satisfy the
conditions\footnote{In principle, we have to introduce two
different exponents $\nu_-$ and $\nu_+$ that will reflect the
behaviors of $f(p)$ near $+\infty$ and $-\infty$; we shall not do
it to lighten the text and just make use of one, denoted by
$\nu$.} (\ref{as1}, \ref{as2}) --- we may check that for $\Lambda$
going to zero we obtain:
\begin{equation}
(\Delta \x)^2_{\Lambda} \propto \left\{
\begin{array}{l}
\Lambda^{\frac{2\nu}{\nu-1}}\rightarrow 0\ \mbox{\rm if}\ \nu<0\\
\Lambda^{\frac{\nu}{\nu-1}}\rightarrow \infty\ \mbox{\rm if}\
0<\nu<1
\end{array}\right.\qquad .
\end{equation}
When $\nu=0$, according to the precise behavior of $f(p)$,
$(\Delta \x)^2_{\Lambda}$ will converge or diverge. If $\nu=1$,
the condition (\ref{as2}) imposes to restrict the domain of
$\Lambda$ to values greater than $1/2$. Moreover in the limit
$\Lambda\to 1/2$, $(\Delta \x)^2_{1/2}$ diverges, while when
$\Lambda$ goes to infinity, as a saddle point estimate shows that
$(\Delta \x)^2_{\Lambda}$ grows as $\Lambda$ and therefore
diverges again. As a consequence, for positive values of $\nu\leq
1$, $(\Delta \x)^2_{\Lambda}$ reaches its minimum for a finite,
non-zero, value of $\Lambda$. This minimum must be strictly
positive, otherwise we obtain also $(\Delta \p)^2_{\Lambda}=0$ and
$\langle f(\p)\rangle_\Lambda=0$, which is of course impossible.
In the special case, $f(p)=1+2\,k\,p+p^{2}$ it is trivial to
obtain the expression of the position uncertainty :
\begin{equation}
(\Delta
\x)^2_{\Lambda}=\frac{\Lambda^{2}}{2\,\Lambda-1}(1+2\,k\,\eta+\eta^{2})
\qquad .
\end{equation}
Minimizing this expression with respect to the variables $\Lambda$
and $\eta$ yields $(\Delta \x)^2 = 1-k^2>0$ for $\Lambda = 1$ and
$\eta = -k$.

However it is clear that the \KMM approach consists in the search
for a possible maximally localized states only among the subset of
states that saturate the Heisenberg inequality (\ref{Hs}): the
squeezed states. But when modified commutation relations are
involved, it is worthwhile to remark that in contrast to what
happens in the framework of usual quantum mechanic, where
$f(\p)=\Id$ and therefore $\langle f(\p)\rangle =1$ whatever is
the state, the expectation value of the operator $f(\p)$, in
general, depends on the state considered. Thus nothing prevents a
minimal length smaller than the one obtained by considering only
squeezed states, simply because there may be physically relevant
states for which $\Delta\x^2$ takes on a lower value than its
lower bound on the subset of squeezed states.

Moreover it is only for a restricted class of functions $f(p)$,
those that do not grow faster than $\vert p\vert^{2}$, that
squeezed states may be used in the chain of relations
(\ref{AA1}-\ref{AA2}). For instance, if we assume $f(p)=1+|p|^\nu$
with $\nu>3$, then all the integrals appearing in eqs
(\ref{dxLdp}) converge for all non negative values of $\Lambda$
and thus the minimal value of $ (\Delta \x)^2_{\Lambda}$ so
estimated is zero. Taking for instance $f(p)=1+p^4$, this seems to
be in contradiction with eq.(\ref{fsp}). But after a rapid glance,
it is clear that there is actually no "contradiction": it just is
that the mean values $\langle\x^2\rangle_\Lambda$ and $\langle
f(\p)\rangle_\Lambda$ do not exist. In the next section we shall
reformulate the problem of the existence of a physical minimal
length on a more appropriate set of states: the subspace of
physical states and show that, indeed, a minimal uncertainty on
the position occurs for such modified commutations relations.
\section{A proposal of physical state space}
As emphasized by \KMM the analysis of the physical implications of
the operators leading to modified commutation relations needs a
careful definition of them. Let us recall that the Hilbert space
we consider is the space of square integrable functions with
respect to the measure $dp/f(p)$. On this space, following \KMM we
start by defining in a first step a symmetric position operator
$\hat x$ given in $p$-representation by $\xp\phi(p)$ with as
domain $\cD({\hat x})$ the set of differentiable functions $\phi$,
vanishing at infinity, and such that
$\int_{-\infty}^{+\infty}f(p)|\partial_p \phi|^2\,dp$ exists. The
extension of this operator to a self-adjoint operator depends on
the behavior of the functions $f(p)$ and $z(p)$ when $\vert
p\vert$ goes to infinity. Two cases have to be considered.
\subsection{The compact case}
Let us suppose that $f(|p|)$ grows at infinity like $|p|^{1+\nu}$,
with $\nu>0$, and therefore that $z(p)$ (eq. (\ref{zu})) goes to
finite limits when $|p|\rightarrow\infty$ :
\begin{equation} z(+\infty)=\alpha_+>0\qquad,\qquad
z(-\infty)=\alpha_-<0\qquad.\label{zfini}
\end{equation} In this case, the mapping $p\mapsto z(p)$ defines a
diffeomorphism between $\RR$ and the interval
$]\alpha_-,\alpha_+[$. The operator $\hat x$ is represented by
$i\,\partial_z$ on this interval and acts on the space
${\cL}^2([\alpha_-,\alpha_+],dz)$ of square integrable functions
vanishing at $\alpha_-$ and $\alpha_+$ and whose derivatives are
also square integrable:
\begin{equation}
\cD(\hat x)=\{\psi(z)\in {\cL}^2([\alpha_-,\alpha_+],dz)\quad{\rm
such\ that}\quad
\psi(\alpha_-)=\psi(\alpha_+)=0\,\}\label{domx}\quad.
\end{equation}
This operator admits a one-parameter family of self-adjoint
extensions \cite{RiNa} $\xl$ whose domains of definition now are
the sets of $\cL^2([\alpha_-,\alpha_+],dz)$ functions with square
integrable derivatives\footnote{By
$\cW^{n,2}([\alpha_-,\alpha_+],dz)$ are denoted the spaces, known
as Sobolev spaces, of square integrable functions whose derivative
are, up to order $n$, also square integrable and whose norms are
$\|f\|^2=\int_{\alpha_-}^{\alpha_+}(|f(z)|^2+\sum_{\ell=1}^n|f^{\ell/}(z)|^2)\,
dz$.} $\cW^{1,2}([\alpha_-,\alpha_+], dz)$ such that:
\begin{equation}
\psi(\alpha_+)=\exp(-i\,\lambda)\psi(\alpha_-)\qquad .
\end{equation} Each of these operators admits an infinite set of
orthonormal eigenvectors:
\begin{eqnarray}
\xl\psi_{n}=\xi_{\lambda,n}\,\psi_{n}&\qquad {\rm
with}\qquad&\psi_{n}(z)=c\
\exp(-i\,\xi_{\lambda,n} z)\quad,\quad\\
\xi_{\lambda,n}=\frac{\lambda+2n\pi}{\alpha_+ -\alpha_-}&\qquad
{\rm and}\qquad& |c|^2=\frac1{\alpha_+ -\alpha_-}\qquad .
\end{eqnarray} If $\nu>2$, these states not only have $\Delta\xl=0$
but also present a finite uncertainty in momentum. Nevertheless
there is no contradiction with the inequality (\ref{AA2}). Indeed,
while the state $\A_\ell\psi_{n}$ is well defined, it did not
belong to the domain of the operator $\Ad_\ell=\A_\ell$; thus we
may not infer the equality (\ref{AA2}).

We also may use states represented in eq. (\ref{sqstate}). They
are always normalizable when $\Lambda>0$ and for all values of
$\Lambda$ if $u(+\infty)$ and $u(-\infty)$ are both finite. In
this last case, the condition (\ref{C1}) is not verified, but even
if $\lim_{p\to\infty}u(|p|)=\infty$ there are the mean values of
the operators $f(\p)=f(p)$ and $\x^2=(i\,f(p)\partial_p)^2$ that
diverge and invalidate the derivation of the equalities
(\ref{AA1}--\ref{AA2}). Contrariwise, if $u(+\infty)$ and
$u(-\infty)$ diverge (for instance if $\nu\leq 1$), $(\Delta
\x)^2$ will be finite and will possibly present a minimum. What we
learn from the previous discussion is that maximally localized
states have not necessarily to belong to the set of squeezed
states, but must be, as least, in the intersection of the domains
of the operators $\x_\lambda$ and $\x^2=-\partial^2_z$. Now comes
into the game a new subtlety: the domain that makes this operator
$\x^2$ self-adjoint can also be defined in different ways. The
most obvious one is to consider $\x^2_\lambda$, whose domain is
given by :
\begin{eqnarray}
\cD(\x^2_\lambda)&=&\cD(\x_\lambda)\cap {\rm
Image}(\x_\lambda)\label{defxl2}\\ &=&\left\{\Psi(z)\in
\cW^{2,2}([\alpha_-,\alpha_+],dz)\ {\rm such\ that\
}\Psi(\alpha_-)=\exp[i\,\lambda]\,\Psi(\alpha_+)\right.\nonumber\\
&&\left. {\rm and}\
\Psi'(\alpha_-)=\exp[i\,\lambda]\,\Psi'(\alpha_+)\right\}\qquad
.\label{bcxl2}
\end{eqnarray}
Let us tentatively define maximally localized states as those
minimizing $\Delta\x^2$ on $\cD(\x^2_\lambda)$:
\begin{equation}\label{x2vp}
(\Delta \x^{min})^2={\rm
min}\frac{\langle\Psi|\x^2_\lambda-\xi^2|\Psi\rangle}{\langle\Psi|\Psi\rangle}\equiv
\mu^2\qquad,\qquad \Psi\in\cD(\x^2_\lambda)
\end{equation}
with
\begin{equation}
\quad\xi
=\frac{\langle\Psi|\xl|\Psi\rangle}{\langle\Psi|\Psi\rangle}\label{locond}
\end{equation}

In $p$-representation this implies that the wave function
$\Psi(p)$ has to obey the equation:
\begin{equation}
\left[-[f(p)\partial_p]^2-\xi^2+2\,a(i\,f(p)\partial_p-\xi)-\mu^2\right]\Psi(p)=0\label{eqpsi}
\end{equation} where $a$ is a multiplicator introduced in order to
take into account the mean position condition imposed on the
state. The parameter $\mu^2$, which corresponds to the value of
the extrema of the functional (\ref{xl2vp}), can also be
interpreted as a multiplicator taking into account the
normalization constraint imposed on the states. The general
solution of eq. (\ref{eqpsi}), satisfying the boundary conditions
(\ref{bcxl2}) reads as:
\begin{equation}
\Psi(p)=C\,\exp\left[i\,\left(\frac{\lambda+2\,n\,\pi}{\alpha_+-\alpha_-}\right)
\,z(p)\right] \qquad,\qquad n\in\ZZ\qquad .
\end{equation} These solutions are simultaneous eigenfunctions of the
operators $\xl$ and $\xl^2$. Thus, in general, they do not satisfy
the localization condition (\ref{locond}), unless $\xi$ is in the
spectrum of $\xl$, in which case $\mu^2$ is zero and there is no
spread in position. The only way we see to cure this defect is to
modify the domain of the operator $\x^2$. Indeed other boundary
conditions than (\ref{bcxl2}) can be chosen, that also makes
$-\partial_z^2$ a self-adjoint operator. We shall define the
squared position operator as the self adjoint operator \cite{RiNa}
$\hat x^\dagger\,\hat x$ whose domain is $\cD(\hat x)$ defined in
eq. (\ref{domx}). The advantage of this choice is that the states
so selected may naturally belong to the domain of some unbounded
operator. More physically, following K. M. M., we shall impose on
the set of states which we consider as physically relevant that
they satisfy an extra condition, for example that they be of
bounded energy. To achieve this requirement we consider an
operator $v(\p)$ such that, in $p$-representation, the
representing function $v(p)$ diverges and is not integrable. We
define the physical states as those whose expectation value of
this operator is finite and reformulate the variational principle
as
\begin{eqnarray} &&(\Delta \x^{min})^2={\rm
min}\frac{\langle\Psi|\x^2-\xi^2|\Psi\rangle}{\langle\Psi|\Psi\rangle}\equiv
\mu^2\qquad, \label{xl2vp}\\ &&{\rm with}\quad\xi
=\frac{\langle\Psi|\xl|\Psi\rangle}{\langle\Psi|\Psi\rangle}\quad,\quad
\gamma=\frac{\langle\Psi|v(\p)|\Psi\rangle}{\langle\Psi|\Psi\rangle}\quad,\quad
\Psi\in\cD(\hat x)\quad .\label{qcond}
\end{eqnarray}
The Euler-Lagrange equation governing this new variational
principle needs the use of one more Lagrange multiplicator and
reads as:
\begin{equation}
\left[-[f(p)\partial_p]^2-\xi^2+2\,a(i\,f(p)\partial_p-\xi)+
2\,b(v(p)-\gamma)-\mu^2\right]\Psi(p)=0\label{eqpsiV}
\end{equation} where $V(z(p))=v(p)$ is the $z$-representation of
$v(\p)$. Of course it is impossible, for an arbitrary function
$v(p)$, to write any exact solution for this equation excepted
when $b=0$. In this case, due to the new boundary conditions, that
have to be taken in order that the mean value of $V(z)$ exists,
the solution we obtain is:
\begin{eqnarray}
&&\Psi(p)=C\,\exp[-i\,\xi\,z(p)]\,\sin\{\mu[z(p)-\alpha_-]\}\\&&|C|=\sqrt{\frac2{\alpha_+-\alpha_-}}
\quad,\quad \mu=n\frac{\pi}{\alpha_+-\alpha_-}\quad,\quad n\in
\NN_0\qquad .\nonumber
\end{eqnarray} What is new with this solution is that it belongs to
$\cD(\hat x)$ and thus is no longer an eigenstate of $\xl$ (but,
of course, belongs also to $\cD(\xl)$). Accordingly, the
corresponding spread in position is non-vanishing but is given by
\begin{equation} \Delta
\x^{min}|_{b=0}=\frac{\pi}{\alpha_+-\alpha_-}\qquad.
\end{equation} But what we have obtained here is the minimal spread
under the extra condition that
\begin{equation}
\frac{\langle\Psi|V(\p)|\Psi\rangle}{\langle\Psi|\Psi\rangle}=\frac2{\alpha_+-\alpha_-}
\int_{\alpha_-}^{\alpha_+}V(z)\sin^2[\frac{\pi(z-\alpha_-)}{\alpha_+-\alpha_-}]\,dz\equiv\gamma_0\quad.
\end{equation} For any other value of $\gamma$, we have, in
principle, to discuss the general solution of eq. (\ref{eqpsiV}),
a task that is virtually impossible. However, the fact that we
obtain a universal solution, ``whatever" is the function $v(p)$
may be the signal that something particular happens when $b=0$.
Indeed, if we vary $\gamma$ by a small amount $\delta\gamma$, a
second order perturbation calculation \cite{LL} leads to
\begin{eqnarray} &&(\Delta \x^{min})^2=(\Delta
\x^{min}|_{b=0})^2+\frac14\frac{\delta\gamma^2}{T^2}\qquad,\qquad
b=-\frac12\frac{\delta\gamma}{T^2}\qquad
\\ &&{\rm where}\qquad
T^2=\frac{2(\alpha_+-\alpha_-)^2}{\pi^2}\sum_{n=2}^\infty\frac{V_{0n}^2}{n^2-1}\\
&& {\rm and}\qquad V_{0n}=
\int_{\alpha_-}^{\alpha_+}V(z)\sin[\frac{\pi(z-\alpha_-)}
{\alpha_+-\alpha_-}]
\sin[n\frac{\pi(z-\alpha_-)}{\alpha_+-\alpha_-}]\,dz
\end{eqnarray}
indicating that $\Delta\x^{min}|_{b=0}$ corresponds to a (local)
minimum with respect to $\gamma$. Such a formal calculation of
course is just indicative, but it allows to conjecture that if we
restrict ourselves to the class of physical states defined as
those belonging to the domain $\cD(\hat x)$ we obtain a minimal
length having a physical meaning. We may also understand this
result in a geometrical way. We have more freedom when we search
for the minimum of $\mu^2$ in $\cD(\hat x)$, without imposing the
value of $\gamma$ than we have for a prescribed value of it.
Therefore the minimum so obtained must be less or equal to those
corresponding to a specified value of $\gamma$. Note also that as
we have assumed {\it a priori} that the solution, belonging to
$\cD(\hat x)$, obtained when $b=0$ defines a finite expectation
value of the operator $v(\p)$, this implies that the function
$v(p)$, defining the physical space, cannot diverge faster than
$|p|^{3\,\nu}$. If this condition is not satisfied we have to
solve eq. (\ref{eqpsiV}) for arbitrary values of $b$, determine
the set of value of $b$ that lead to finite expectation values of
$v(\p)$ and then minimize the values of $\Delta\x_b$ with respect
to $b$.
\subsection{The non-compact case}
From the discussion of the previous subsection it is clear that if
$\alpha_-$ or $\alpha_+$ become infinite, {\it i.e.} if $\nu<0$
the squeezed states define, for $\Lambda>0$, physical states (with
respect to any polynomial operator $v(\p)$) whose spread in
position can be made as small as we want. Therefore there is no
minimal length in this case. The semi bounded case (corresponding
to one of the $\alpha$ finite and the other not) will not be
considered here.
\section{Two physical illustrations}
The previous analysis may appear to be formal and far from our
physical apprehension of localization in ($x$)-space. So, to
convince the reader that our approach is well founded, let us
briefly discuss the quantization of a particle in a box
\cite{BGLS} and the asymptotic behavior of the energy eigenvalues
of the harmonic oscillator \cite{K2}.

Of course, we only have to consider the case that we have called
compact, the one where the function $z(p)$ tends to finite value
when $|p|$ goes to infinity. Let us introduce the inverse of this
function, a function $\zeta(z)$ such that
\begin{equation}
\zeta[z(p)]\equiv p\qquad .
\end{equation}
Note that this function becomes singular in $\alpha_-$ and
$\alpha_+$. In $x$-space we may represent (formally) the $\x$ and
$\p$ operators as:
\begin{equation}
\x = x\qquad,\qquad \p=\zeta(-i\partial_x)\qquad .
\end{equation}
For the problem of quantizing a particle of mass $m$ in a box of
length $L$, we may take as domain the square integrable,
$C^\infty$, $\cL^2([0,L],\, dx)$ functions, vanishing at $x=0$ and
$x=L$. To simplify the discussion, let us first assume that the
function $f(p)$ is even and therefore that $\zeta$ is odd. We have
to solve the eigenvalue equation:
\begin{equation}\label{zbox}
\p^2\Psi=2\,m\,E\,\Psi \qquad \Leftrightarrow\qquad
\zeta(-i\partial_x)^2\psi(x)=2\,m\,E\,\psi(x)
\end{equation}
whose solutions are
\begin{equation}
\psi(x)\propto \sin(k\,x)\quad {\rm with}\quad k\,L=n\,\pi\ ,\
n\in\NN_0\quad{\rm and}\quad 2\,m\,E=\zeta(k)^2\qquad .
\end{equation}
Imposing that the energy is bounded and is a monotone function of
the number of nodes of the wave function, we deduce the existence
of an upper bound $N_{max}$ on the number of nodes and therefore
of the dimensionality of the Hilbert space of eigenstates:
\begin{equation}
N_{max}=\left[\alpha_+\frac L\pi\right]
\end{equation}
where, as usual, the brackets represent the function "integer
part". This illustrates the existence of a minimal length in the
theory: distances less than $\pi/\alpha_+$ are not
distinguishable. If the function $\zeta$ is not odd,the eigenvalue
problem (\ref{zbox}) is solved as follows. For definiteness, let
us assume $\alpha_+<\alpha_-$. We divide the interval
$[0,\alpha_+]$ into subintervals of length $\pi/L$. In each of
such subinterval we may find a value of $k_+$ such that there
exists a value of $k_-$ obeying the condition \begin{equation}
k_-=-k_+2\,n\,\frac \pi L\quad n\in \ZZ\quad{\rm so\ that
}\quad\zeta(k_+)=-\zeta(-k_-)\qquad.
\end{equation}
The solutions of eq. (\ref{zbox}) are therefore of the form:
\begin{equation}
\psi(x)\propto \exp(i\,k_+x)-\exp(-i\,k_-x)
\end{equation}
\begin{equation}
{\rm with}\ k_+L\in [(n-1)\,\pi,\ n\pi]\ ,\ 1\leq n< \alpha_+\frac
\pi L\quad{\rm and}\ 2\,m\,E=\zeta(k_+)^2=\zeta(k_-)^2\ .\nonumber
\end{equation}
The monotonic behavior of the function $\zeta$ ($d\zeta(z)/dz
=f[\zeta(z)]$) insures the existence and uniqueness of these
solutions for each value of $n$.

We also may easily obtain the asymptotic behavior of the energy
eigenvalues of an harmonic oscillator. The eigenvalue problem to
be solved is
\begin{equation}
\left(\frac{\p^2}{2\,m}+\frac{m\,\omega^2}2\x^2\right)\Psi_n
=E_n\Psi_n\qquad .
\end{equation}
In term of the $z$ variable this problem is formally equivalent to
the usual problem of a particle of mass $1/m\,\omega^2$, in a
potential that diverges like $|\alpha_\pm-z|^{\nu_\pm}$. For large
value of the energy $E_n$ we may approximate this as the motion of
a free particle in a box of size $\alpha_+-\alpha_-$. Accordingly,
the asymptotic eigenvalues of the energy are given by:
\begin{equation}
E_n\approx\frac{m\,\omega^2}2(\frac
\pi{\alpha_+-\alpha_-}n)^2\qquad,
\end{equation}
in accordance with the special case explicitly integrated in ref.
\cite{K2}. The physical interpretation of this result \cite{RB} is
the following. When the wave function, confined in an $x^2$
potential, presents a large number $n$ of nodes its spatial
extension is of the order on $n\, |\Delta \x^{min}|=n\, \frac
\pi{\alpha_+-\alpha_-}$ and its energy is, for large values of
$n$, approximated by
\begin{equation}
E_n\approx\frac{m\,\omega^2}2(n\, |\Delta \x^{min}|)^2
\end{equation}
contrary to the standard case where $\Delta \x =\sqrt{\x^2}\approx
n/\Delta\p|$ and $\Delta\p^2=\p^2\approx m\,E_n$, leading to
$E_n\approx n\,\omega$ for large value of $n$.

These examples already appear in the works \cite{BGLS, K2},
assuming $f(p)=1+p^2$. In this case, as we have emphasized in this
article, one may also find maximal localization using squeezed
states. Our point here is that one may derive maximal localization
from analogous studies of the wave equation with more general
functions $f(p)$, only subject to the conditions (\ref{zfini}), in
which case these maximally localized states are not squeezed
states, but they can be derived from the variational principle
exhibited in (\ref{xl2vp}, \ref{qcond}). Furthermore the above
techniques applied to the asymptotic part of the spectrum of the
wave equation is clearly generalizable to any potential function
which is sufficiently slowly varying in $x$ space.
\section{Maximally localized states in $n$-dimension }
The way to extend the commutation relation (2) to $n$ dimensions
is not unique. Let us now assume that $\x$ and $\p$ (resp. $\bf x$
and $\bf p$) are vector-valued operators (resp. vectors) of
components $\xj$ and $\pk$ (resp. ${\bf x}_j$ and ${\bf p}_k$)
that obey the commutation relations
\begin{equation} [\x_j ,\pk] =
i \left[f(\p ^{\,2})\delta_{jk}+g(\p^{\,2}) \pj \pk \right] \equiv
i \hat \Theta_{jk} \label{xpnd} ,
\end{equation}
which ensure both translation and rotation invariance. Requiring
the position operator components to satisfy $[\x_j ,\xk]=0$
further imposes the constraint (see \cite{AK}]
\begin{equation}
g = \frac{2 f \partial_{p^2} f}{f-2 p^2 \partial_{p^2}f} \quad
.\label{tc}
\end{equation}
Similarly to what has been done in one dimension where $z(p)$
acted on the space ${\cL}^2([\alpha_-,\alpha_+],dz)$ as an
operator such that $[\x,\z] = i$, we introduce the operator $\z$
whose components are canonically conjugated to the components of
the position operator :
\begin{equation}
[\xj, \zk] = i \, \delta_{jk} \qquad , \label{xznd}
\end{equation}
represented by functions ${\bf z}_j ({\bf p})$ in
$p$-representation. We can therefore, by representing $\pj$ as the
functions $\zeta_j({\bf z})$ in (\ref{xpnd}), rewrite this
commutation relation as
\begin{equation}
\partial_{z_j}{\bf p}_k = i \Theta_{jk} \qquad .\label{A}
\end{equation}
Thus $\partial_{z_j}{\bf p}_k=-\partial_{z_k}{\bf p}_j $ and
therefore there exists a scalar function $\cU$ generalizing
$[u(p)]$ (\ref{zu}) such that
\begin{equation}
p_j = \partial_{z_j} \cU \qquad . \label{B}
\end{equation}
We shall assume this function $\cU$ to depend only on the variable
${\bf z}^2/2$ in order to preserve rotation invariance. Denoting
the derivative with respect to $\bz^2 /2$ as
\begin{equation}
\frac {\partial \cU}{\partial (\bz^2 /2)}=\cU ' \qquad ,
\end{equation}
we deduce from (\ref{A}) and (\ref{B}) that
\begin{equation}
f({\bp}) = {\cU}'[\bz(\bp)]\quad \et\quad g(\bp) = \frac{
{\cU}''[\bz(\bp)]} { {\cU}'[\bz(\bp)]} \qquad .
\end{equation}
%\fin
Under the assumption that relation (\ref{B}) can be inverted, we
can, in principle, express $|{\bf z}|$ as a function of $|{\bf
p}|$: $|{\bf z}| = \cZ (|{\bf p}|)$ and we may check that equation
(\ref{tc}) is indeed verified. Furthermore eq. (\ref{B}) defines a
diffeomorphism between $\mathbb{R}^n$ and the $n$-dimensional ball
:
\begin{equation}
\cB^n = \{ {\bf z} \in \mathbb{R}^n \quad \mbox{such as} \quad
|{\bf z}| < \rho_+ \}
\end{equation}
where $\cZ ( +\infty) = \rho_+ < +\infty$ in the compact case, the
only one we consider here after.

A useful representation of the algebra displayed in (\ref{xpnd})
and (\ref{xznd}), is on the Hilbert space ${\cL}^2(\cB^n,d^nz)$,
as:
\begin{eqnarray}
\x_j \psi ({\bf z}) &=& i \partial_{z_j} \psi ({\bf z}) \nonumber\\
\zj \psi ({\bf z}) &=& z_j \psi ({\bf z}) \nonumber\\
\pj \psi ({\bf z}) &=& z_j {\cU}' (\frac{{\bf z}^2}{2}) \psi ({\bf
z})
\end{eqnarray}
with $\psi ({\bf z}) = \langle {\bf z} \vert \Psi \rangle$. \KMM
suggested to build maximally localized states by considering the
solution of the $n$ equations
\begin{eqnarray}
\A_{j \Lambda}|\Psi_{\Lambda}\rangle\equiv\left[(\xj-\langle
\xj\rangle)+i\Lambda(\pj-\langle\pj\rangle)
\right)|\Psi_{\Lambda}\rangle=0\qquad \label{CSn}
\end{eqnarray}
where $\Lambda$ has to be independent of the index $j$ for the
equations to be consistent. Their common solution reads as
\begin{equation}
\psi ({\bf z}) \propto \exp[-i \langle\x\rangle {\bf z} + \Lambda
\langle\p\rangle {\bf z} - \Lambda \cU ({\bf z}^2/2)] \quad .
\end{equation}
But as we already discussed in the one-dimensional framework, this
construction, excepted in a few special cases, does not lead to
states with maximal localization or can even not be performed when
no restriction ensuring that the operators $\A_{i}$ and $\A_{j}$
commute for all $i$ and $j$ is imposed on the commutation
relations. In that case, we may decide to improve localization in
some direction at the cost to deteriorate localization in another
direction. As we illustrate below for the particular case $g(\p
^2) =1$ considered in \cite{KeMa} where the squeezed states
construction can be achieved, those states {\it do not} display
the minimal length uncertainty. Here again, our purpose will be to
define maximally localized states as those minimizing $(\Delta
\x)^2$ with $\langle \x \rangle = {\xi}$ fixed and verifying
suitable boundary conditions in order to be physically acceptable.
Therefore, following the same procedure as in sect. ({\bf 3}), we
find that the wave function reads as
\begin{equation}
\psi ({\bf z}) = \exp[-i \, {\xi}.{\bf z}] \phi ({\bf z}) \quad
\mbox{where} \quad ({\bf \Delta} + \mu ^2)\phi ({\bf z}) = 0, \,\,
\phi ({\bf z})_{\vert \partial \cB^n} =0 \, ,
\end{equation}
$\xi$ denoting now the vector mean position and ${\bf \Delta}$ the
$n$-dimensional Laplacian operator on $\cB^n$. The boundary
condition $\psi ({\bf z})_{(|{\bf z}|=\rho_+)} = 0$ leads to the
quantization of $\mu ^2$, whose smallest value is obtained for the
$s$-wave. The corresponding wave function, bounded at the origin,
is
\begin{equation}
\psi ({\bf z}) = C_n \,\exp[-i\,\xi. {\bf z}] \,\vert{\bf
z}\vert^{-\nu}\, J_{\nu} (\mu |{\bf z}|)\qquad {\rm where}\qquad
\nu=\frac n2 - 1
\end{equation}
displaying the finite minimal uncertainty in position
\begin{equation}
\Delta \x^{min} = \frac{j_{\nu,1}}{\rho_+} \qquad, \label{xmin}
\end{equation}
$j_{\nu,1}$ being the first zero of the $J_{\nu} (x)$ Bessel
function.

We end up this section by comparing, for illustrative purpose, the
spread in position of the squeezed states obtained by \KMM in the
particular case for which
\begin{equation}
{\cU}' (\frac{{\bf z}^2}{2}) = \frac{1}{1-{\bf z}^2/2}
\end{equation}
with the spread (\ref{xmin}). The normalized squeezed states with
$\langle\x\rangle = {\bf 0}$ and $\langle\p\rangle = {\bf 0}$ read
as
\begin{equation}
\psi_{\Lambda} ({\bf z}) = \left[(2 \pi)^{-n/2} \frac{\Gamma(2
\Lambda + \frac{n}{2} +1)}{\Gamma(2 \Lambda + 1)}\right]^{1/2}
\left(1- \frac{{\bf z}^2}{2}\right)^{\Lambda} \quad .
\end{equation}
The squared uncertainty in position computed from these states, as
a function of $\Lambda$, is
\begin{equation}
(\Delta {\bf x}^2) = \frac{n \Lambda}{4} \frac{n + 4 \Lambda}{2
\Lambda -1}
\end{equation}
whose minimum is
\begin{equation}
(\Delta \x^{min}_{squeezed})^2 =\frac{n}{8} \left(1 + \sqrt{1 +
\frac{n}{2}}\right) \left(\frac{n+2}{\sqrt{1 + \frac{n}{2}}} + 2
\right) \,\, \mbox{for} \,\, \Lambda = \frac{1}{2} +
\frac{1}{2}\sqrt{1 + \frac{n}{2}} \label{xsq}
\end{equation}
is always greater than the one given by eq. (\ref{xmin}).
Incidentally, let us notice that from eqs (\ref{xmin}) and
(\ref{xsq}) we obtain the inequality
\begin{equation} \label{maJ}
j_{\nu,1}\,< \sqrt{\left(\nu+\frac12\right)\left(\nu+2\sqrt{\nu +
\frac 32}+\frac 52\right)}
\end{equation}
which provides an upper bound to the first non-trivial zero of
$J_{\nu} (x)$ when $2\,\nu +1$ is an integer.
\section{Conclusions}
We have shown the emergence of a minimal length in quantum
mechanics starting from very general assumptions on the
commutation relations. We have clarified the meaning of maximally
localized states, using to build them the most natural definition
we may consider: a variational principle minimizing the spread of
the state in space. We have also shown that the \KMM construction
of such states in terms of squeezed states in general fails to
provide maximally localized states. A particularly interesting
physical consequence of our discussion is that in the regime where
the minimal length becomes sensitive, the physics seems to become
independent of the details of the modification of the commutation
relations and depends only on the convergence of the integral
$\int_{-\infty}^{+\infty}f(p)^{-1}\,dp$, as our analysis of the
free particle and the harmonic oscillator examplifies. Our
definition of maximally localized states extends also, without
difficulties, to any number of dimensions. As a mathematical
by-product of our discussion, we obtain a large class of
majoration of the first zero of a family of Bessel functions.
Indeed for each choice of the function $\cU$ we may obtain a
relation like eq. (\ref{maJ}), the one displayed in the text being
better than the Schafheitlin's one (for the values of the index
$\nu$ considered here) \cite{Wa}.
\section{Acknowledgments}
We thank A. Kempf for many valuable comments on a preliminary
version of this paper and for having drawn our attention on one of
its previous work in which a similar variational principle was
considered to determine the state of lowest momentum uncertainty
for a particle in a box (see sect. 2.1 of ref. \cite{K6}). We are
also particular grateful to M. Lubo for stimulating discussions
and comments. Moreover, one of us (Ph.S.) would like to thanks its
colleagues R. Brout, C.~Demol and C. Troestler for enlightening
conversations. Ph.S. acknowledges support from the Belgian Fonds
National de la Recherche Scientifique (FRFC contracts).
\appendix
\section{Appendix: an {\it a priori} minoration of $\Delta\x$} If
$f(\p)$ can be expressed as a series of even power of $\p$, with
positive coefficients:
\begin{equation} f(\p)=1+\sum_{n=1}^\infty c_n\,\p^{2n}\qquad {\rm
with}\qquad c_n\geq0\qquad.
\end{equation} we obtain
\begin{equation}
\langle f(\p)\rangle\geq 1+\sum_{n=1}^\infty
c_n\,\langle\p^2\rangle^n \qquad.
\end{equation} The proof of this inequality leads on the positivity of
\begin{equation}
\left\langle(\p^2-\langle\p^2\rangle)^2\,\p^{2n-4}\right\rangle\geq
0
\end{equation} from which we deduce that
\begin{equation}
\langle\p^{2n}\rangle\geq
2\langle\p^2\rangle\langle\p^{2n-2}\rangle -
\langle\p^2\rangle^2\langle\p^{2n-4}\rangle
\end{equation} and by recurrence that
\begin{equation}
\langle\p^{2n}\rangle \geq \langle\p^2\rangle^n\qquad.
\end{equation} As a consequence, when formula (\ref{dxLdp}) is valid,
we deduce a lower bound for the squared position uncertainty on
squeezed states :
\begin{equation} (\Delta\x)^2\geq(\Delta\x)^2_{min}=\frac 14 {\rm
min}\frac{f^2(\sqrt{\langle\p^2\rangle})}{\langle\p^2\rangle}\qquad
.\label{fsp}
\end{equation} 
\end{document}